\documentclass[]{spie}  

\usepackage{amsmath,amsfonts,amssymb}
\usepackage{graphicx}
\usepackage[colorlinks=true, allcolors=blue]{hyperref}
\usepackage{comment}
\usepackage{caption}
\usepackage{mathrsfs}
\usepackage{upgreek}
\usepackage{float}
\usepackage{ulem} 

\usepackage{color,soul}

\newcommand{\MM}[1]{\textcolor{black}{#1}}

\newcommand{\ML}[1]{\textcolor{black}{#1}}

\newcommand{\red}[1]{\textcolor{black}{#1}}

\title{Phase noise characterisation of a 2-km Hollow-Core Nested Antiresonant Nodeless Fibre for Twin-Field Quantum Key Distribution }

\author[a]{M. Minder*}
\author[a]{S. Albosh*}
\author[b]{O. Alia}
\author[c]{R. Slavik}
\author[a]{R. Kumar}
\author[c]{F. Poletti}
\author[b]{G. Kanellos}
\author[a]{M. Lucamarini}
\affil[a]{School of Physics, Engineering \& Technology and York Centre for Quantum Technologies, Institute for Safe Autonomy, University of York, York, YO10 5FT, UK}
\affil[b]{ High Performance Network Group, School of Computer Science, Electrical \& Electronic Engineering and Engineering Maths (SCEEM), University of Bristol BS8 1TH, Bristol, UK}
\affil[c]{Optoelectronic Research Centre, University of Southampton, Southampton, SO17 1BJ, UK}

\authorinfo{*Equal contribution to this work \\
Correspondence to:\\
M.M. (mariella.minder@york.ac.uk), S.A. (sophie.albosh@york.ac.uk), M.L. (marco.lucamarini@york.ac.uk)}

\date{11 January 2023}
\begin{document} 
\maketitle
\begin{abstract}
The performance of quantum key distribution (QKD) is heavily dependent on the physical properties of the channel over which it is executed. 
Propagation losses and perturbations in the encoded photons' degrees of freedom, such as polarisation or phase, limit both the QKD range and key rate. 
The maintenance of phase coherence over optical fibres has lately received considerable attention as it enables QKD over long distances, e.g., through phase-based protocols like Twin-Field (TF) QKD\cite{Lucamarini2018}. 
While optical single mode fibres (SMFs) are the current standard type of fibre, recent hollow core 
fibres (HCFs)\cite{8535324, 9083240, 9605918} could become a superior alternative in the future. 
Whereas the co-existence of quantum and classical signals in HCF has already been demonstrated\cite{9605918}, the phase noise resilience required for phase-based QKD protocols is yet to be established. 
This work explores the behaviour of HCF with respect to phase noise for the purpose of TF-QKD-like protocols. 
To achieve this, two experiments are performed. 
The first, is a set of concurrent measurements on 2~km of HCF and SMF in a double asymmetric Mach-Zehnder interferometer configuration. 
The second, uses a TF-QKD interferometer consisting of HCF and SMF channels. 
These initial results indicate that HCF is suitable for use in TF-QKD and other phase-based QKD protocols.
\end{abstract}


\section{INTRODUCTION}
\label{sec:intro}  
Quantum key distribution (QKD) is a technology that can provide information-theoretic security in the distillation of a key between distant users.
To achieve this, the users must encode information in the quantum states of single photons and exchange them over a channel, such as an optical fibre. 
Due to the loss of photons during their transmission, the maximum distance over which QKD can be performed is limited\cite{Pirandola2017}. 
With the recently developed Twin-Field QKD\cite{Lucamarini2018} (TF-QKD) protocol, this distance has been doubled, with implementations reaching 830~km\cite{Wang2022}. 
In addition to losses, any further perturbations in the channel affecting the single photons' degrees of freedom will also decrease the secret key rate and maximum distance achievable with QKD, as they introduce errors between the prepared and measured states. 
For example, in some point-to-point QKD systems, it is critical that random polarisation rotations caused by the quantum channel are compensated for prior to the detection of the single photons. For TF-QKD, in addition to polarisation, phase coherence of the single photon pulses must be maintained, for as long as hundreds of kilometres of optical fibre. This entails challenging requirements for phase stabilisation.

Single-mode fibre (SMF) is the most commonly used channel for the transmission of terrestrial optical communications and therefore QKD systems are often designed to operate via these channels. However, recent developments in hollow core fibres (HCFs)\cite{8535324, 9083240, 9605918}, and in particular in \red{nested anti-resonant nodeless HCFs (NANF)}, suggest that they  can offer intrinsic benefits to the channel's performance in comparison to SMFs. 

\begin{figure}
\captionsetup{width=1.0\linewidth}
\centering 
\includegraphics[width=0.85\linewidth]{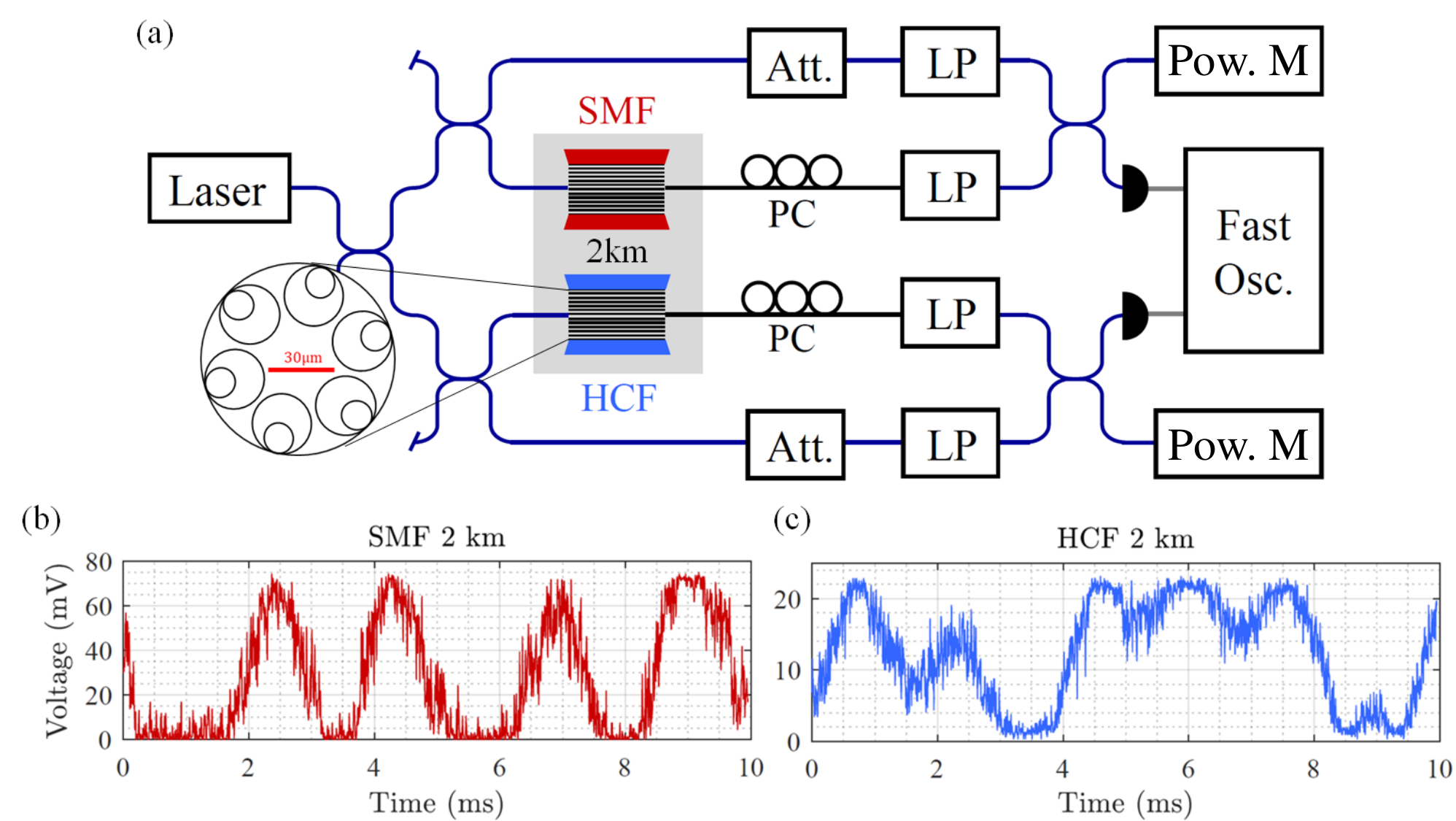}\\[3mm]
\caption{\label{Fig_AMZI_setup_trace}
\textbf{Schematics of the experimental setup.} 
(a) Parallel AMZIs built side by side for testing a \red{nested anti-resonant nodeless hollow core fibre (NANF), indicated as `HCF' in the figure and in the text,} in comparison with an SMF. The core structure of the \red{HCF} is represented in the expanded diagram.
The setup consists of a Rio Orion laser module, 50:50 polarisation maintaining beam splitters, two 2~km spools, one for the \red{HCF} and one for the SMF (Thorlabs SMF-28-1000), variable optical attenuators (Att.), polarisation controls (PC), linear polarisers (LP), power meters (Pow. M), fast photodetectors and a 36 GHz oscilloscope (Fast Osc.). 
(b) and (c) traces of the photodetector voltage at the output of the \red{HCF} and SMF AMZIs, recorded on the fast oscilloscope with a sampling frequency of 200~kS/s.}
\end{figure}

HCFs \red{guide light through a hollow core, 
therefore inheriting} characteristics from both \red{optical} fibres and free-space channels. 
In terms of loss, HCF have reached attenuation coefficients similar to that of standard SMF\cite{9083240}, but it is estimated that they have the capacity to achieve record low attenuation values, surpassing state-of-the-art solid core fibres\cite{9748272}. 
In addition to this, HCF also exhibits other desirable properties for the purposes of QKD, including lower latency, reduced chromatic-dispersion and nonlinearity. 
For protocols other than TF-QKD, investigations into the suitability of HCF in QKD have been explored, for example the co-existence of classical signals with photons in the Coherent One Way protocol has been shown\cite{9605918}. BT have also performed trials involving QKD and HCF\cite{BT_HC-NANF_QKD_2021}.
However, the characterisation of a HCF in a typical TF-QKD setup has not been performed so far. 
The relevant question in this regard is whether the larger-than-SMF HCF core (see diagram in Fig.~\ref{Fig_AMZI_setup_trace}(a)) and the presence of membranes within it, necessary to confine and guide the light through the fibre, result in HCF picking up noise different from the SMF that spoils the phase coherence needed in TF-QKD.

Here, we answer this question by performing two experiments using two separate 2-km spools of HCF and SMF in parallel, see Fig.~\ref{Fig_AMZI_setup_trace}(a).
The first experiment, is a set of concurrent measurements on HCF and SMF in a double asymmetric Mach-Zehnder interferometer (AMZI) configuration, where the resulting interference data is used to assess the phase noise across a wide spectral range. The second experiment, involves the construction of a TF-QKD-like setup with HCF and SMF channels, where the results provide direct insight into the suitability of HCF for the purposes of TF-QKD.

\section{METHODS and Results}
\label{sec:methods}
In this work a 2~km sample of \red{HCF} is used, which contains six nested tubes\cite{8535324} (see diagram in Fig.~\ref{Fig_AMZI_setup_trace}(a)) and is spooled around a bobbin with a diameter of $\sim 30$~cm. The HCF sample is spliced to short FC/APC connectorised SMF patch cables at either end, where these splices contribute significantly to the optical loss of the fibre; approximately $6$~dB. 
In addition to this, a 2~km spool of SMF (Thorlabs SMF-28-1000) with a standard attenuation coefficient of $0.18$~dB/km at 1550~nm\cite{THORLABS_SMF} is used as a control experiment. 

\subsection{Phase Noise Investigation}
To perform an initial assessment of the phase properties of the \red{HCF}, we compare the HCF and the SMF directly in a double-AMZI configuration. 
The rationale is that SMFs have been effectively used for TF-QKD.
Therefore, a similar noise figure from the HCF would suggest its suitability for use in a TF-QKD transmission. 

To this purpose, two AMZIs are constructed as described above and
as shown in Fig.~\ref{Fig_AMZI_setup_trace}(a). 
Continuous wave (CW) laser light from a narrow linewidth ($<1$~kHz) Rio Orion laser module is sent into the two interferometers simultaneously. To improve interference visibility, manual variable optical attenuators are used to balance the loss between the arms of the AMZIs and linear polarisers are used to refine the polarisation state of the light just prior to interference. Polarisation maintaining fibre and components are used where possible, however the HCF and SMF fibre spools are not polarisation maintaining, and therefore polarisation controls are used to optimise the throughput of the linear polarisers. The CW laser light is detected using power meters and fast 5~GHz photodetectors. The power meters are employed to monitor the optical power during setup and optimisation of the polarisation controls and the variable optical attenuators. The voltage outputs of the photodetectors are recorded on a 36~GHz oscilloscope, which allows for fast phase fluctuations experienced by the CW laser light to be resolved.

Voltage traces are recorded for three different sampling frequency and measurement duration combinations: 200~kS/s for 20~s, 5~MS/s for 5~s and 500~MS/s for 0.1~s. A section of the voltage traces for the SMF and HCF AMZIs are shown in Figs.~\ref{Fig_AMZI_setup_trace}(b) and ~\ref{Fig_AMZI_setup_trace}(c), respectively, for a sampling frequency of 200~kS/s. By performing a moving average of this data, conservative estimates for the interference visibility can be deduced, giving $>99\%$ for the SMF and $>92\%$ for the HCF. These values are limited by imperfect attenuation balance between the interferometer arms and due to the fact that they are acquired with classical detectors, which are prone to dark current and fluctuating DC bias. Typically, visibility estimates $>90\%$ attained using classical detectors are indicative of actual visibility values $>95\%$ when measured with single-photon detectors, which is confirmed in the next section. 

\begin{figure}[t]
\noindent\makebox[\textwidth][c]{%
\begin{minipage}[c]{1\textwidth}
(a) \hspace{8cm} (b) \\[1mm]
$\vcenter{\hbox{\includegraphics[width=0.5\linewidth]{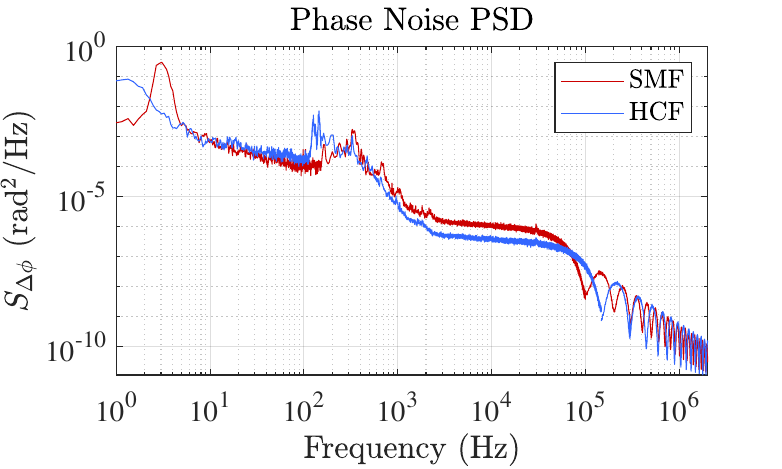} }}$
$\vcenter{\hbox{\includegraphics[width=0.5\linewidth]{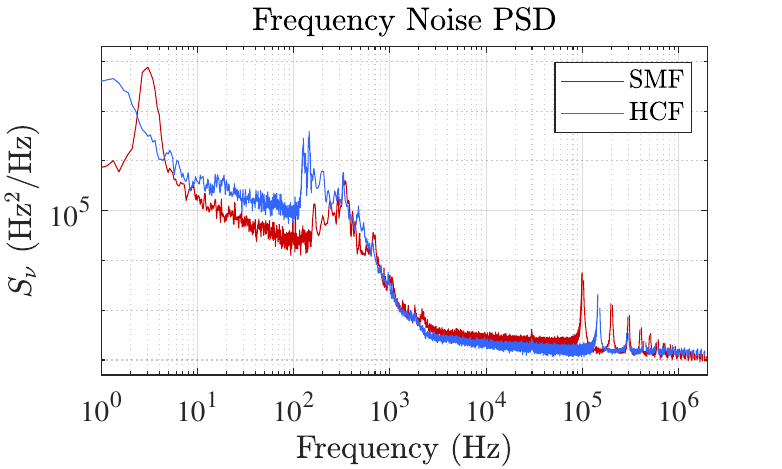} }}$ \vspace{0.3cm}
\captionsetup{width=1.0\linewidth}
\caption{
\label{Fig_PSD} 
Experimental PSD characterisation.
(a) The phase noise PSDs of the SMF and HCF AMZIs, calculated using Welch's method and a Hann window with a width $1\times 10^6$ samples. 
(b) The frequency noise PSDs of the SMF and HCF AMZIs calculated from the phase noise PSDs.
}
\end{minipage}}
\end{figure}

The voltage traces are converted into phase values bounded between $(0,\pi)$, which are in turn used to calculate the phase noise power spectral density (PSD), $S_{\Delta\phi}$ (see appendix~\ref{sec:PSD_calcs} for details). 
The PSDs are formulated using Welch's method\cite{1161901}, with a Hann window of $1\times 10^6$ samples and $50\%$ overlap between windows.
This is inspired by similar analysis conducted in previous TF-QKD work\cite{Clivati2022}. 
The resulting phase noise PSDs for the HCF and the SMF AMZIs are shown in Fig.~\ref{Fig_PSD}(a). 
They are constructed from the three sampling frequency and measurement duration sets, enabling spectral analysis that extends from $1$~Hz to $2$~MHz. 
The characteristic ripples above $50$~kHz are a result of the temporal asymmetry between the two arms of the AMZIs. 
To accurately measure these delays, the CW laser light is modulated to produce narrow pulses. The SMF and HCF AMZIs have temporal asymmetries of 9.9~$\upmu$s and 6.6~$\upmu$s respectively, owing to the different refractive indices of the silica and air cores. 
This well matches the periodicity of the ripples in the plots.

Generally speaking, the PSD plots sit above the PSD derived from the laser phase noise. 
We verified this feature in a separate experiment using a standard laser phase noise
characterisation technique\cite{4738475}.
More specifically, we can identify three regions of phase noise in the PSDs graphs.

The first is \red{between $1$~Hz and $10$~Hz, a region that typically is dominated by thermal fluctuations}. 
In this range a peak is evident in the SMF PSD plot at 3~Hz, \red{which we attribute to} changes in the optical path length of the SMF due to the thermo-optic effect and thermal expansion.
As the SMF spools are exposed to varying surrounding temperatures, the optical path difference between the arms of the AMZI will change, where the large asymmetry of the interferometer results in a constant phase shift, producing sinusoidal interference patterns throughout the measurement duration. 
This effect is less \red{pronounced} in the HCF PSD \red{and is visible at a lower frequency, below 2~Hz}, \red{an indirect evidence of} the proven reduced thermal sensitivity of HCF compared to SMF\cite{Slavik2015}. 
All the low-rate phase fluctuations do not represent an obstacle in QKD and can be easily corrected using phase stabilisation techniques.

The second region of phase noise is between $10$~Hz and $1$~kHz \red{and we attribute it mainly to acoustic or mechanical vibrations.}
The mechanism by which vibrations impact the phase is difficult to model, \ML{even more so because of the different physical spooling dimensions of HCF and SMF. 
However, we managed to pinpoint several} correlations between acoustic frequencies in the lab and features in the phase noise PSDs.
\ML{Moreover, irrespective of the explicit identification of the sources of noise, it is important to note that the HCF PSD is remarkably close to that of the SMF.
As SMFs and lasers, with a comparable phase noise figure to the one shown here, have already been used in TF-QKD with excellent results\cite{MPR+19,PML21}, we argue that even at these frequencies the HCF would perform equally well for TF-QKD.}

\ML{In the region above $1$~kHz, correcting phase drift becomes technically} more challenging, \ML{but still possible, because of the reduced amplitude in the PSD graph.} 
\ML{At these frequencies,} the laser phase noise can be seen, in particular the small peak at 30~kHz, which has been confirmed during the experimental process. 
\ML{To make the comparison between the HCF and SMF PSDs more straightforward, we converted} the phase noise PSD, $S_{\Delta\phi}$, into a frequency noise PSD, $S_\nu$, removing the offset between the two plots caused by the different optical path lengths of the two fibre samples, \ML{due to having equal-length fibres with different index of refraction}. 
The frequency noise PSD is calculated using the methods outlined in appendix \ref{sec:PSD_calcs} and the plots are shown in Fig.~\ref{Fig_PSD}(b). 
For \ML{frequencies} greater than 1~kHz, \red{we expect comparable noise for HCF and SMF, mainly contributed by the laser phase noise. 
This is confirmed by the results in Fig.~\ref{Fig_PSD}(b), where the difference in amplitude between the two PSDs is minimal and the ripples are due to the different index of refraction in the two fibres.}

\subsection{Proof-of-concept TF-QKD experiment}


\noindent \ML{In a second experiment, we reconfigured} the setup to perform a proof-of-concept TF-QKD demonstration with the HCF.
\ML{To our knowledge, this is the first evidence of experimental TF-QKD over HCFs.}

\begin{figure}
\captionsetup{width=1.0\linewidth}
\centering 
\includegraphics[width=1\linewidth]{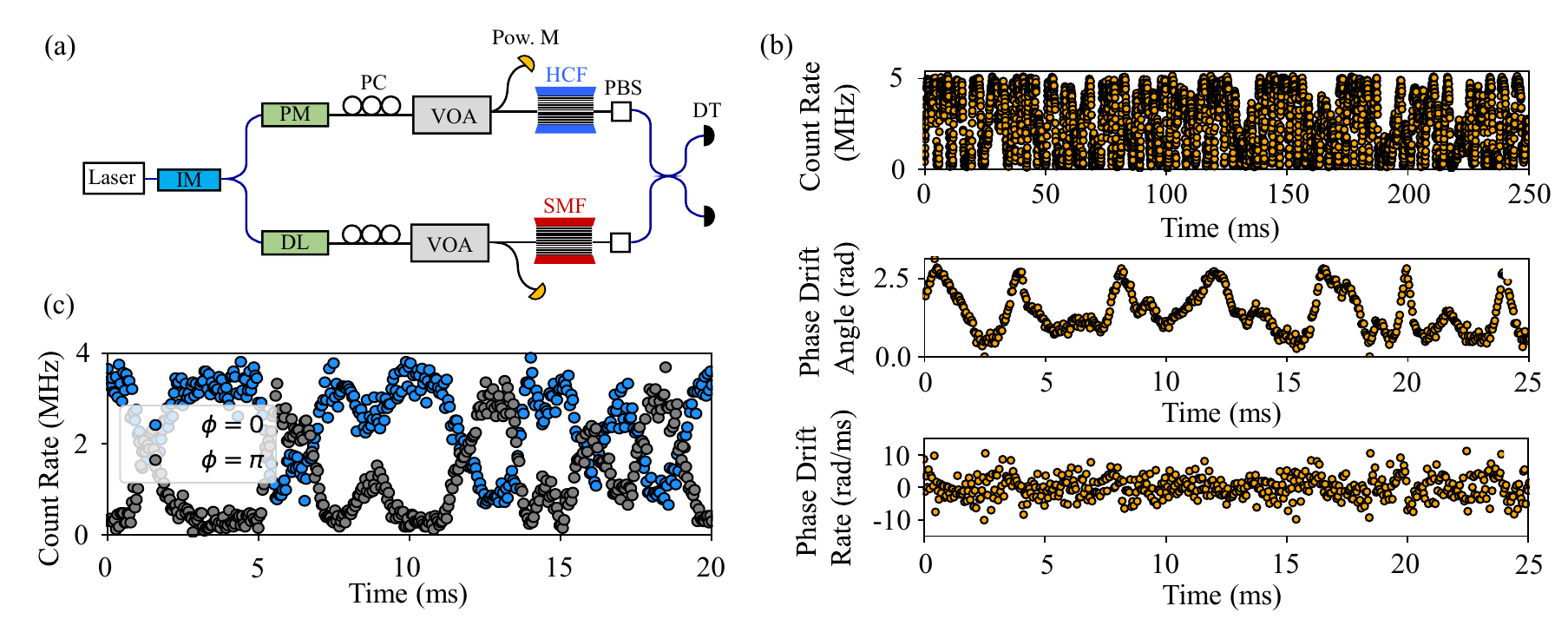}\\[3mm]
\caption{\label{Fig:qkdSetup} 
\ML{\textbf{TF-QKD setup and interference.}}
(a) The test TF-QKD experimental setup, consisting of CW laser, intensity modulator (IM), a phase modulator (PM) \ML{on the HCF path and a delay line (DL) on the SMF path}, polarisation controls (PC), variable optical attenuators (VOA), power meters (Pow. M), polarisation beam splitters (PBS) and \ML{detectors (DT).}
(b) The count rate recorded over 250~ms on the SNSPD when no phase modulation is applied (upper panel); 
\ML{the derived} phase drift (middle panel) and phase drift rate (lower panel) over \ML{the first} 25~ms \ML{of the sample}.
(c) The count rate recorded over 20~ms when 0 and $\pi$ phase modulation is applied on the PM on the HCF arm.
}
\end{figure}

The experimental setup is shown in Fig~\ref{Fig:qkdSetup}(a), comprising an AMZI where the interferometer arms are constructed using the 2~km HCF and SMF samples. 
Pulsed light is generated using an intensity modulator and CW light from a Rio Grande high power laser module.
Phase shifts are applied to the optical pulses propagating through the HCF arm using a phase modulator and an optical delay line is used to temporally overlap the pulses for interference at the final beam splitter.
Polarisation controls are used to optimise throughput of the polarisation beam splitters and the loss between the AMZI arms is balanced using variable optical attenuators.
For optimisation of the setup, bright pulses are generated and  detected using classical photodetectors.
For the TF-QKD demonstration, the pulses are attenuated to the single-photon level and monitored using an IDQuantique ID281 superconducting nanowire single photon detector (SNSPD).

To assess the phase drift rate of the setup, initially no phase modulation is applied to the interferometer and the resulting count rates of the SNSPD are recorded and shown in \ML{the top panel of} Fig~\ref{Fig:qkdSetup}(b).
By taking the maximum and minimum count rate, \ML{we calculate} the corresponding phase of the signal \ML{attenuated to the single-photon level}. 
From this, \ML{we determine} the phase drift angle 
and the phase drift rate, both of which are also shown in Fig~\ref{Fig:qkdSetup}(b).
The standard deviation of the phase drift rate is $7$~rad/ms, with instantaneous phase drifts of up to 10~rad/ms, resulting from a combination of the phase noise \ML{sources identified in the previous section} impacting both the HCF and SMF samples in the laboratory environment.
This phase drift rate is significant but not uncommon for asymmetric implementations of TF-QKD and can be compensated via phase stabilisation techniques\cite{Clivati2022}.
\red{The implementation presents asymmetric optical paths due to the different refractive indices of the two 2-km fibres, which causes a fast drift in phase fluctuations, faster than typical symmetrical implementations\cite{Lucamarini2018}, due to the residual phase noise of the laser.} 
We anticipate that using two equal length samples of HCF in each arm of the TF-QKD setup would provide a much lower phase drift rate than evaluated in this setup.

We repeated the tests on pulsed light, generated through the high-extinction ratio ($\sim$~60dB) intensity modulator (IM) inserted after the laser, see Fig.~\ref{Fig:qkdSetup}(a).
To resemble a real TF-QKD setup, we used a repetition rate of 1~GHz with \MM{200~ps} pulses and aligned the interference at the final beam splitter using an optical delay line (DL).
The resulting interference visibility is in excess of 99\%.
In this configuration, we tested the capability of the setup to implement the basic feature of a TF-QKD modulation. 
In the coing mode of the TF-QKD protocol described in Ref.~\citen{CAL19} the users modulate their pulses between only two phase values, $0$ and $\pi$. 
We reproduced this pattern here using the phase modulator in Fig~\ref{Fig:qkdSetup}(a).
We recorded the \MM{time-tagged counts} of the SNSPDs and sorted them into the phase modulated sets. 
The resulting drift over 20~ms is shown in Fig~\ref{Fig:qkdSetup}(c).
Despite the fast drift of the phase, the $0-\pi$ modulation pattern is recognisable from the data and provides a quantum bit error rate (QBER) of 1.75\%, which is in line with previous TF-QKD implementations\cite{WYH22}. 
It is worth mentioning that this value represents the base QBER of the system as it includes both the errors due to the free drift of the phase and due to an imperfect modulation.

\section{Conclusion}

We performed a set of experiments to assess the suitability of a novel 2~km nested antiresonant nodeless hollow core fibre (NANF) for phase-based quantum key distribution (QKD) protocols, in particular twin-field (TF) QKD. 
In so doing, we provided the first proof-of-concept implementation of TF-QKD over HCF on a kilometre scale. 
Our first set of results show that HCF features a remarkable resistance to phase noise, not only in the low-frequency regime\cite{MFH22}, but also in the high-frequency spectrum of interest to TF-QKD. 
After normalising the effect of the asymmetry in the optical paths, the HCF displays a comparable or better phase noise performance than SMF.
Because SMF has been successfully used in TF-QKD, our findings suggest that HCF is also suitable for TF-QKD and for other quantum protocols that are sensitive to the phase degree of freedom.
We confirmed this result with a second test performed on a TF-QKD-like interferometer, in the presence of 1~GHz pulse carving and phase modulation, showing the high visibility of the resulting interference fringes.

\red{The HCF robustness to phase fluctuations adds positively to an already conspicuous set of features that are advantageous to QKD. 
The excellent polarization stability of the HCF simplifies the alignment of the distant users' reference systems required in QKD. 
Moreover, the extremely low nonlinearity of the HCF makes it possible to multiplex the quantum signals with the bright pulses needed in the classical communication of QKD or in the phase correction of TF-QKD, without imparting the QBER of the transmission. 
This promotes HCFs as an ideal medium for implementing QKD and TF-QKD in particular} and confirms the central role they can play in next-generation quantum  communications.

\appendix

\section{Calculation of the power spectral density} 
\label{sec:PSD_calcs}

%
\ML{Here, we provide explicit calculations for the PSD presented in the main text, in line with the ones in Ref.~\citen{Clivati2022}.}

The power at the output of the AMZI is given by,
\begin{equation}
P = \frac{1}{2} \left\{ S + D\cos [ \Delta \phi(t) ] \right\} \label{EQN_Int_Phase}
\end{equation}
where $P$ is the power detected at the fast detector,
$S$ and $D$ are the sum and difference of the maximum and minimum of the interference fringe, $S = \max(P) + \min(P)$ and $D = \max(P) - \min(P)$, and $\Delta\phi(t)$ is the phase difference between the arms of the interferometer. The phase noise PSDs shown in Fig.~\ref{Fig_PSD}(a) are calculated using the values of $\Delta\phi$ directly, performing Welch's method with a Hann window of width $1\times 10^6$ samples and a window overlap of $50\%$. For an AMZI with delay $\tau$, the phase difference can be expressed as,
\begin{equation}
\Delta\phi(t) = \phi(t+\tau)-\phi(t).
\end{equation}
The Fourier transform of this phase results in the following,
\begin{align}
\Delta\Phi(f)   & = \mathscr{F}\left(\Delta\phi(t)\right) \nonumber\\
                &= \int^{\infty}_{-\infty} \Delta\phi(t) e^{-2\pi i ft} dt \nonumber\\
                &= \left( 1 - e^{-2\pi i f \tau} \right)\Phi(f) 
\end{align}
where $\Phi(f) = \mathscr{F}\left(\phi(t)\right)$. The phase PSD is proportional to the absolute square of the Fourier transform, therefore the equivalent phase PSD relation is given by,
\begin{equation}
S_{\Delta\phi}  = \left| 1 - e^{-2\pi i f \tau} \right|^2 S_{\phi}  =  4\sin^2\left( \pi f\tau \right)S_{\phi}
\end{equation}
where $S_{\Delta\phi}$ is the phase noise PSD calculated from the phase change $\Delta\phi$. The phase PSD term, $S_{\phi}$, contains contributions to the phase noise from the laser and the fibre spools, but does not include the features due to the AMZI temporal asymmetry, $\tau$. The frequency PSD in Fig.~\ref{Fig_PSD}(b) is calculated using the following expression,
\begin{equation}
S_{\nu}  = f^2 S_{\phi}.
\end{equation}

\acknowledgments 
Funding has been provided through the partnership resource scheme of the EPSRC Quantum Communications Hub grant (EP/T001011/1).

\bibliography{main} 
\bibliographystyle{spiebib} 

\end{document}